\documentclass[12pt,a4paper]{article}
\usepackage[utf8]{inputenc}
\usepackage{authblk}
\usepackage{subfiles}
\usepackage{tabularx}
\usepackage{booktabs} 
\usepackage{caption}  
\usepackage{array}    
\usepackage{float}
\usepackage{setspace}
\usepackage{siunitx}
\usepackage[font=footnotesize,labelfont=bf]{caption}
\usepackage{xcolor}
\usepackage{soul}
\sethlcolor{yellow}
\sethlcolor{yellow!30}
\usepackage{soulutf8}
\usepackage{setspace}
\usepackage{helvet}
\usepackage[left=3cm, right=2.5cm]{geometry}
\usepackage{graphicx}
\usepackage{enumitem}

\usepackage{subfiles}
\usepackage{gensymb}
\usepackage{lscape}
\usepackage[normalem]{ulem}
\useunder{\uline}{\ul}{}
\usepackage{amsmath}
\usepackage{mathtools}
\usepackage{comment}
\usepackage{amsfonts}
\usepackage{hyperref}
\usepackage{ragged2e}
\usepackage{nomencl}
\usepackage{indentfirst}

\makenomenclature

\hypersetup{
    colorlinks=true,
    linkcolor=black,
    filecolor=blue,      
    urlcolor=blue,
    citecolor=blue
    }
\makeatother

\graphicspath{ {./images/} }
\title{\LARGE{Tunable Electrocaloric Effect in Lead Scandium Tantalate through Calcium Doping}}
\author[1,2]{\normalsize{Youri Nouchokgwe*}}
\author[1,2]{Natalya S. Fedorova}
\author[1,2]{Veronika Kovacova}
\author[2,3]{Pranab Biswas}
\author[4]{Ivana Gorican}
\author[4]{Nejc Suban}
\author[4]{Silvo Drnovsek}
\author[4]{Matej Sadl}
\author[3]{Michele Melchiorre}
\author[1,2]{Binayak Mukherjee}
\author[1,2]{Uros Prah}
\author[5]{Guillaume F. Nataf}
\author[1,2]{Torsten Granzow}
\author[2,3]{Mael Guennou}
\author[4]{Hana Ursic}
\author[1,2,3]{Jorge Iñiguez-González}
\author[1,2,3]{Emmanuel Defay*}
\affil[1]{\footnotesize{\textit{Smart Materials Research Unit, Luxembourg Institute of Science and Technology, Maison des Matériaux, 28 Av. des Hauts Fourneaux, L-4362 Esch-Belval, Esch-sur-Alzette.}}}
\affil[2]{\footnotesize{\textit{Inter-institutional Research Group University of Luxembourg - LIST on Ferroic Materials,41 rue du Brill, L-4422 Belvaux, Luxembourg}}}
\affil[3]{\footnotesize{\textit{Department of Physics and Materials Science, University of Luxembourg, L-4422 Belvaux, Luxembourg}}}
\affil[4]{\footnotesize{\textit{Jožef Stefan Institute, Jamova cesta 39, Ljubljana 1000, Slovenia}}}
\affil[5]{\footnotesize{\textit{GREMAN UMR7347, CNRS, University of Tours, INSA Centre Val de Loire, 37000 Tours, France}}}
\affil[*]{\footnotesize{\textit{\underline{corresponding authors:} youri.nouchokgwe@list.lu ; emmanuel.defay@list.lu}}}

\begin{document}

\maketitle

\begin{abstract}

State-of-the-art electrocaloric cooling prototypes rely on the conventional electrocaloric effect of ferroelectric lead scandium tantalate (PbSc$_{0.5}$Ta$_{0.5}$O$_{3}$, PST), which peaks near room temperature. Here, we demonstrate that A-site calcium doping in highly ordered PST modifies its phase transitions and enables precise tuning of the electrocaloric response. The transition temperature shifts down to 258\,K and up to 319\,K, depending on Ca concentration. Calorimetry under electric field, electrical polarization loops, and piezoresponse force microscopy reveal the emergence of an intermediate antiferroelectric phase stabilized for Ca $\geq$ 2\%. These results are supported by first-principles calculations. We observe conventional electrocaloric effect for Ca $\leq$ 2\% and inverse electrocaloric effect at higher doping ($\geq$ 2\%). Under an applied field of 110\,kV\,cm$^{-1}$, Ca-doped PST exhibits an adiabatic temperature change of 2\,K over a range from 263\,K to 353\,K. Such Ca-doped PST compounds could be used to expand the temperature range of PST below the freezing point of water. Our results offer a pathway to cascaded electrocaloric cooling devices with extended operating spans.  

\end{abstract}

\section{Introduction}

Electrocaloric (EC) cooling technology has been proposed as an energy-efficient \cite{moya2015too, defay2018enhanced} and environmentally friendly alternative to conventional vapor-compression systems. Demonstrations of EC prototypes \cite{Torello2022} have shown achievable temperature spans of about 20\,K, typically between 293\,K and 313\,K \cite{li2023high}. The main refrigerant in EC prototypes is an electrocaloric material that heats up upon the application of an electric field and cools down when the field is removed. This phenomenon, known as the conventional EC effect, reaches its maximum near the ferroelectric phase transition, where the polarization varies most sharply with temperature \cite{Moya2014, Moya2020}. State-of-the-art EC prototypes exploit $\sim 2$\,K electrocaloric response \cite{li2023high} of lead scandium tantalate (PbSc$_{0.5}$Ta$_{0.5}$O$_{3}$, PST) associated with its first-order ferroelectric to paraelectric transition near room temperature. However, PST exhibits a negligible electrocaloric effect \cite{Nair2019, nouchokgwe2022scale} at lower temperatures, limiting its use for applications below ambient conditions.

PST is a dielectric material with a perovskite-type A(B'B'')O$_{3}$ structure, where the B-site is occupied by both scandium (Sc) and tantalum (Ta) cations \cite{Setter1981}. It is considered a benchmark material for studying the EC effect \cite{Nair2019, nouchokgwe2025quantifying}. The degree of cation ordering at the B-site, quantified by the long-range order parameter $\Omega$, plays a crucial role in determining the amplitude of the EC effect. A higher degree of B-site ordering enhances the EC effect \cite{Nouchokgwe2021}. The parameter $\Omega$ ranges from 0 (completely disordered) to 1 (perfectly ordered). In a perfectly ordered PST, Sc and Ta cations alternate in adjacent B-sites in a 1:1 fashion, leading to a double perovskite unit cell. High degrees of B-site ordering are typically achieved through long-duration annealing treatments at elevated temperatures \cite{Stenger1979, Stenger1980, stenger21980}. PST undergoes a displacive phase transition from a rhombohedral structure of \(R{3}\) or \(R{3}m\) symmetry at low temperatures to a cubic structure (space group \(Fm\bar{3}m\)) at higher temperatures \cite{Stenger1980, stenger21980, Baba-Kishi1990}. This transition has a spontaneous electric polarization associated with it, emerging from the displacement of the Pb and B-site cations along a $<$111$>$ crystallographic direction \cite{Baba-Kishi1990}. The phase transition of PST is influenced by the degree of ordering. Disordered PST exhibits relaxor behaviour \cite{Granzow2025PST}, characterized by a broad phase transition over a wide temperature range, as observed in dielectric measurements by Setter et al. \cite{Setter1981, Setter1980}. However, as the B-site cation ordering increases, PST becomes more like a conventional ferroelectric (FE), and a sharp phase transition from a polar FE phase at low temperatures to a non-polar paraelectric (PE) phase at high temperatures occurs. Similarly, the transition temperature of PST depends on the degree of ordering. It ranges from 260\,K to 310\,K, with highly ordered PST exhibiting higher transition temperatures \cite{nouchokgwe2022scale}.

Phase transformations in both disordered and ordered PST ceramics and single crystals have been extensively studied. An intermediate phase between the FE and PE phases has been reported in the literature \cite{isupov2003ferroelectric}. In the 1990s, Baba-Kishi et al. \cite{Baba-Kishi1990} identified an incommensurate antiferroelectric (IAFE) phase in PST using transmission electron microscopy. The authors observed satellite reflections along the [100] directions in the reciprocal space, which were believed to originate from the inequivalent antiparallel displacement of Sc and Ta cations, leading to the formation of an AFE phase. The incommensurate nature of this phase was indicated by the fact that no rational indices could be assigned to the satellite reflections. This contrasts with a commensurate transition, where polarization is periodically modulated as a rational multiple of the lattice parameter. The IAFE phase was observed between 233\,K and 313\,K, spanning a temperature range of 80\,K, where the FE and PE phases coexist. Two decades later, Dul'kin et al. \cite{Dulkin2010} investigated the phase transformations in disordered PST using acoustic emission. They detected two acoustic emission signals at 261\,K and 293\,K, which were attributed to the transitions from FE to IAFE and IAFE to PE, respectively. These studies suggest that PST contains antiferroelectric (AFE) regions, indicating the coexistence of FE, IAFE, and PE phases in the material. However, no direct evidence, such as electrical double hysteresis loops, has yet been provided to conclusively demonstrate the presence or stability of an AFE phase in PST. 

In this study, we investigate the phase diagram and EC effect of highly ordered PST doped on the A-site with calcium (Pb$_{1-x}$Ca$_x$Sc$_{0.5}$Ta$_{0.5}$O$_3$). Highly ordered PST ceramics with four different calcium contents (0, 1, 2, and 4.6 mol\%) were prepared. The estimated B-site ordering of these samples ranges between 0.88 and 0.92. Polarization–electric field loops reveal that A-site calcium doping above 1 mol\% stabilizes an AFE phase between the FE and PE phases. This intermediate AFE phase becomes increasingly stable with higher calcium content. These results are corroborated by piezoresponse force microscopy, in situ field calorimetry, dielectric measurements and density functional theory calculations. Furthermore, Ca doping strongly affects the transition temperature of PST, shifting it down to 258\,K and up to 319\,K, depending on composition. Using an infrared camera, we measured the EC response under applied electric fields in Ca-doped PST compounds. Conventional EC effect was observed for Ca contents $\leq 2\%$, while inverse EC effect appeared for Ca contents $\geq 2\%$. For the ferroelectric compounds PST and 1\%Ca-PST, maximum adiabatic temperature changes ($\Delta T_{\mathrm{adiab}}$) of 4.6~K at 313~K and 2.6~K at 285~K, respectively, were obtained at their FE--to--PE phase transitions. In 2\%Ca-doped PST, which exhibits two successive phase transitions (FE--to--AFE and AFE--to--PE), both conventional and inverse EC effects were measured. At an applied electric field of 115~kV~cm$^{-1}$, maximum $\Delta T_{\mathrm{adiab}}$ values of 2.0~K and 2.5~K were obtained at the FE--to--AFE and AFE--to--PE transitions, respectively. At low applied electric fields ($< 30$~kV~cm$^{-1}$), a negligible inverse EC effect of $-0.25$~K was observed at the AFE--to--PE transition. The antiferroelectric 4.6\%Ca-PST compound exhibits a pronounced inverse EC effect at its AFE--to--PE transition, reaching a maximum $\Delta T_{\mathrm{adiab}}$ of $-0.6$~K at 303~K. With increasing electric field and temperature, this inverse effect progressively transforms into a conventional EC effect, with a maximum $\Delta T_{\mathrm{adiab}}$ of 1.8~K at 323~K.
These findings demonstrate that Ca-doped PST (Ca $\leq$ 2\%) are promising candidates for cascaded electrocaloric cooling devices with enlarged temperature spans.    

\section{Results}

\subsection{Structure of Ca-doped PST}

Four compositions of PST were fabricated and investigated: pure PST ceramic, along with three PST ceramics doped on the A-site with calcium at concentrations of 1, 2, and 4.6 mol\%. All samples were prepared by mechanochemical synthesis and heat treatment as described in the Methods section. For clarity, Ca-doped PST is abbreviated as PCaxST, where x refers to the calcium concentration. Specifically, PCa1ST, PCa2ST, and PCa4.6ST represent PST doped with 1, 2, and 4.6 mol\% calcium, respectively. We measured an average density of 8696\,kg\,m$^{-3}$ across all samples (see Table \ref{tablePCaxST}) corresponding to 96\% of the theoretical density for PST \cite{Liu2010}.

Scanning electron microscope (SEM) images were taken at room temperature to assess the grain size of the ceramics (see Supplementary Note 1). As detailed in Table \ref{tablePCaxST}, the average grain sizes of all samples are similar, namely in the range between 1.1 and 1.7$\mu$m. The introduction of calcium did not significantly affect the grain size. SEM images (Supplementary Figure 1) confirm the absence of secondary phases in all samples. X-ray diffraction (XRD) was performed on crushed ceramics (see Supplementary Figure 3). Measurements were taken at 323\,K to ensure that all samples were in the same cubic phase. The B-site cation ordering parameter, $\Omega$, was calculated based on the ratio of integrated intensities $I$ of the XRD peaks at 111 and 200, as defined in \cite{Shebanov1989}. $\Omega$ is determined by taking the square root of the ratio of the experimental $I_{111}$/$I_{200}$ to the theoretical $I_{111}$/$I_{200}$. The latter was calculated from the simulated XRD pattern of a supercell structure (see Methods section and Supplementary Figure 28) with perfect alternation of Sc and Ta ($\Omega$ = 1). As shown in Table \ref{tablePCaxST} and Supplementary Table 4, the theoretical $I_{111}$/$I_{200}$ increases with Ca content. The experimentally obtained $\Omega$ values range between 0.88 and 0.92, indicating the high level of order in our ceramics and showing the stability of $\Omega$ with respect to doping.

\subsection{Electrical measurements of Ca-doped PST}

\begin{figure}[!ht]
    \centering
    \vspace{3ex}%
    \includegraphics[width=1.1\textwidth]{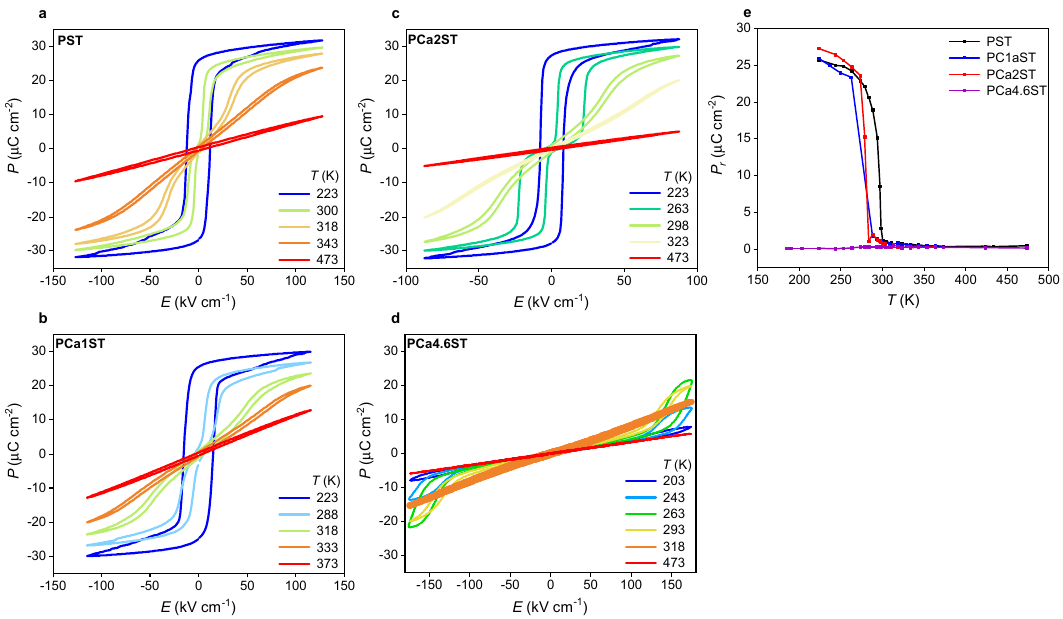}
    \caption{\textbf{Polarization – Electric field loops of Ca-doped PST.} (a), (b), (c), and (d) show the polarization (\textit{P}) versus electric field (\textit{E}) loops of respectively PST, PCa1ST, PCa2ST, and PCa4.6ST at different temperatures \textit{T}. These \textit{P}-\textit{E} loops were taken on heating. (e) Remanent polarization (\textit{P}$_r$) of the same four ceramic compositions as a function of temperature (\textit{T}).}
    \label{peloops}
\end{figure}

An Aixacct tool (see Methods section) was used to study the temperature dependence of the polarization versus electric field (\textit{P-E}) loops of the four ceramic compositions. For clarity, \textit{P-E} loops at selected temperatures are displayed in Figure \ref{peloops}. Supplementary Note 3 shows the loops at all the temperatures studied. As expected, pure PST (Figure \ref{peloops}a) exhibits a typical FE hysteresis loop at low temperatures. PST transitions to a PE phase around 300\,K. This PE phase is marked by a linear PE loop at high temperatures. PCa1ST shows a transition from a FE loop to a PE loop (see Figure \ref{peloops}b). However, the FE to PE transition in PCa1ST occurs at approximately 288\,K, which is lower than the transition temperature observed in PST. For PCa2ST, two distinct transitions are observed (see Figure \ref{peloops}c). The first transition, occurring at approximately 263\,K, is indicated by a change from a pronounced FE loop to a characteristic double hysteresis AFE loop. The second transition, from the AFE to a PE loop, is observed around 298\,K. 
Lastly, Figure \ref{peloops}d demonstrates that PCa4.6ST exhibits AFE behaviour down to 184\,K, with a transition to a PE phase at 318\,K. 
Figure \ref{peloops}e describes the remanent polarization ($P_r$) of PCaxST samples as function of temperature $T$. 
PCa4.6ST exhibits nearly zero remanent polarization within the studied temperature range (184 K to 473 K), and no FE hysteresis loop was detected. However, PST, PCa1ST, and PCa2ST display a sharp increase in $P_r$ at lower temperatures, indicating the presence of a FE phase in these materials. It is worth mentioning that dielectric measurements (Supplementary Note 4) confirm similar transition temperatures in the Ca-doped PST compounds with low loss tangent of 0.040.

\subsection{Thermal analysis of Ca-doped PST}

\begin{figure}[!ht]
    \centering
    \vspace{3ex}%
    \includegraphics[width=1.1\textwidth]{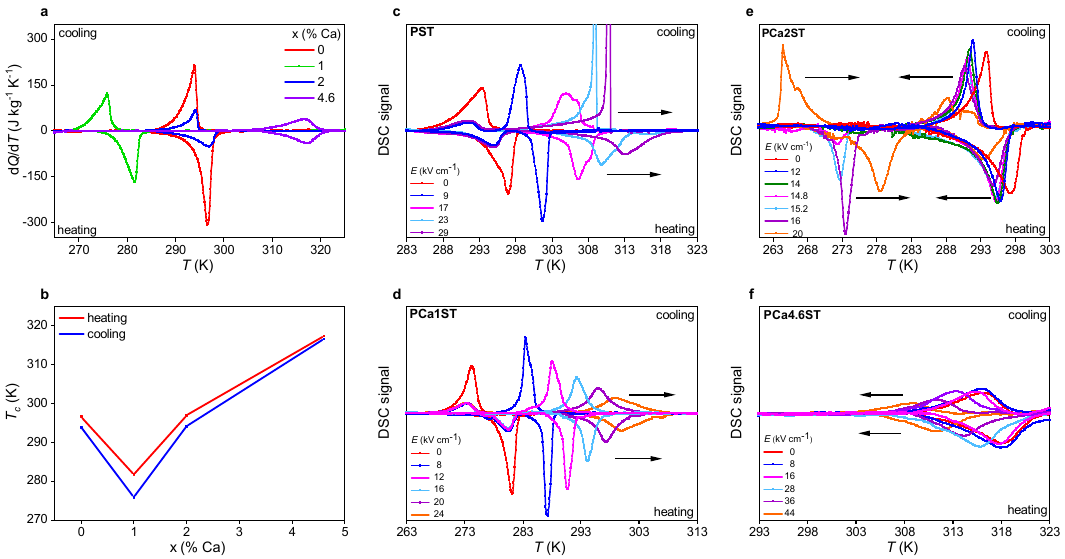}
    \caption{\textbf{Thermal analysis of Ca-doped PST.} (a) Heat flow (d\textit{Q}/d\textit{T}) of PST, PCa1ST, PCa2ST, and PCa4.6ST on cooling (top) and heating (bottom) under no applied electric field. (b) The Curie temperature (\textit{T$_c$}) of PCaxST as a function of the calcium content x. (c), (d), (e), and (f) show the isofield DSC signal of respectively PST, PCa1ST, PCa2ST, and PCa4.6ST at different constant electric field $E$. The top and bottom curves are the heat flow when cooling and heating, respectively. The black arrows indicate the movement of the peaks under the constant electric field $E$.}
    \label{thermalanalysis}
\end{figure}

Figure \ref{thermalanalysis} shows the zero-field heat flow and isofield measurements of the four PCaxST compositions carried out using a customised differential scanning calorimeter (DSC). These measurements were performed with a low heating/cooling rate of 5\,K\,min$^{-1}$ (see Methods section), to obtain accurate position of the transition temperatures. 

Sharp peaks, latent heat (integral under the peaks) and thermal hysteresis are observed in the zero-field heat flow data (Figure \ref{thermalanalysis}a). These three features confirm the first-order transition behaviour of our ceramics. It is noted that the sharpness of the peaks decreases with the calcium content, yet still significant to be considered as first-order phase transition materials. In pure PST, we measure a Curie temperature ($T_c$) of 297\,K on heating and 294\,K on cooling (Figure \ref{thermalanalysis}b). This describes a transition from a FE phase to a PE phase. PCa1ST also transitions from a FE to a PE phase, however, at lower temperatures compared to PST (see Figures \ref{thermalanalysis}a and \ref{thermalanalysis}b). Indeed, $T_c$ moves from 297\,K at 0\%Ca to 282\,K at 1\%Ca on heating, and from 294K at 0\%Ca to 276\,K at 1\%Ca on cooling. The Curie temperature of PCa1ST measured by DSC on heating (282\,K) is close to the transition temperature determined from the \textit{P-E} loops (287\,K). PCa2ST shows a Curie temperature of 297\,K on heating and 294\,K on cooling. These corresponds, respectively, to the AFE to PE and PE to AFE phase transitions. The AFE-to-PE transition temperature (297\,K) matches the transition temperature (298\,K) obtained from the polarization measurements (Figure \ref{peloops}c). It is worth noting that the FE-to-AFE transition in PCa2ST observed in the \textit{P–E} loops is not visible in the zero-field heat flow.  The Curie temperature of PCa4.6ST is shifted to higher temperatures by 20 K upon heating and by 23 K upon cooling relative to undoped PST (Figure \ref{thermalanalysis}b). PCa4.6ST exhibits a transition from the AFE to the PE phase. The transition temperatures measured here coincide with those observed in the $P–E$ loops (see Figure~\ref{peloops}d). 

We now apply a constant electric field in our Ca-doped PST and sweep in temperature (isofield measurements). In the ferroelectric compound PST (Figure \ref{thermalanalysis}c), upon application of an electric field, the transition peak shifts toward higher temperatures, as indicated by the black arrows. Similar behaviour occurs in the ferroelectric compound PCa1ST (see Figure \ref{thermalanalysis}d). Indeed, in both compounds, the low-temperature FE phase is stabilized under an electric field. Notably, in both ferroelectric compounds, a remnant of the zero-field peak persists under applied field, albeit with reduced intensity (Figures \ref{thermalanalysis}c and \ref{thermalanalysis}d). This residual peak corresponds to the non-electroded region of the sample, where no electric field is applied. 

In PCa2ST, under electric fields ranging from 12\,kV\,cm$^{-1}$ to 20\,kV\,cm$^{-1}$, the AFE-to-PE transition peak (297\,K) slightly shifts to lower temperatures (Figure \ref{thermalanalysis}e). At the electric field of 14.8\,kV\,cm$^{-1}$, a new peak appears at 272\,K, corresponding to the transition from the FE to the AFE phase. This transition, which was not visible at electric fields smaller than 14.8\,kV\,cm$^{-1}$ is now shifted to higher temperatures with increasing field, stabilizing the FE phase. At the same time, the AFE-to-PE transition slightly shifts to lower temperatures, gradually suppressing the AFE phase with increasing field.  

Isofield measurements in the antiferroelectric compound PCa4.6ST (Figure \ref{thermalanalysis}f) show that the AFE-to-PE phase transition clearly shifts to lower temperatures as the electric field increases.

\subsection{Phase diagram of Ca-doped PST}

\begin{figure}[!ht]
    \centering
    \vspace{3ex}%
    \includegraphics[width=1.1\textwidth]{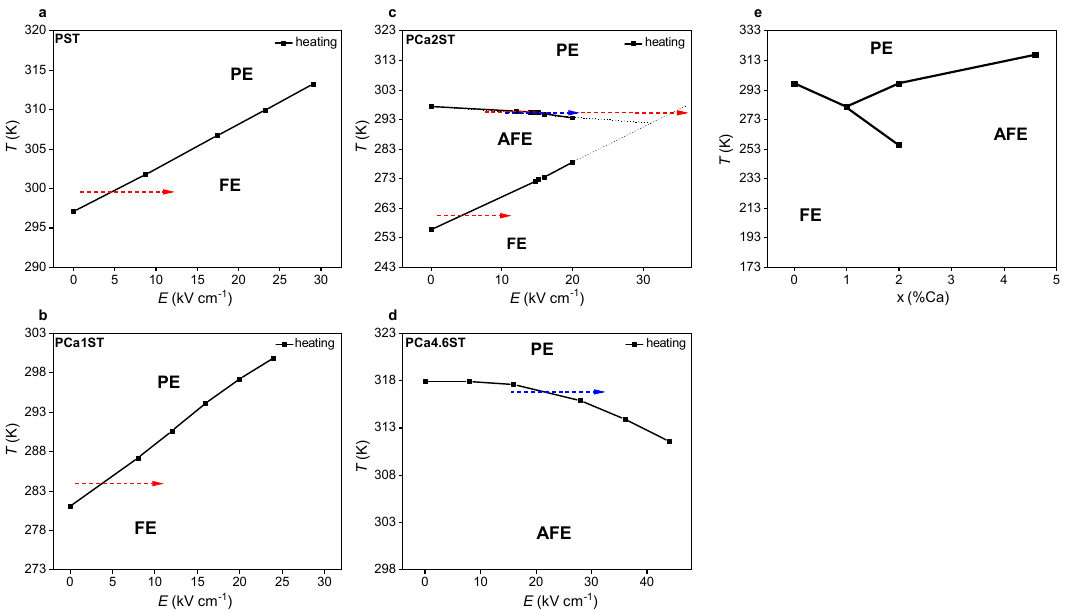}
    \caption{\textbf{Phase diagrams of Ca-doped PST.} FE, AFE and PE are respectively the ferroelectric, antiferroelectric and paraelectric phases. (a), (b), (c), and (d) show the phase diagrams (transition temperature $T$ as a function of electric field $E$) of respectively PST, PCa1ST, PCa2ST, and PCa4.6ST. The data were collected from the isofield DSC measurements (see Figure \ref{thermalanalysis}). For clarity, the data shown here correspond to heating. The arrows indicate the effect of applying an electric field near the phase transitions. Their colours represent the anticipated electrocaloric response: red for a conventional electrocaloric effect and blue for an inverse electrocaloric effect. (e) Transition temperature and phases of Ca-doped PST as a function of the calcium concentration x. }
    \label{phasediagram}
\end{figure}

From the heat flow measurements (Figure \ref{thermalanalysis}), we constructed, in Figure \ref{phasediagram}, the phase diagrams ($T$–$E$) of the corresponding PCaxST compositions, illustrating the transition temperature $T$ as a function of the applied electric field $E$.

The phase diagrams of PST and PCa1ST (Figures \ref{phasediagram}a and \ref{phasediagram}b, respectively) show a ferroelectric–paraelectric (FE–PE) phase sequence. Their FE phase (at low temperatures) becomes increasingly stabilized with higher electric fields. Moreover, the dependence of their transition temperature on the electric field ($\frac{dT}{dE}$) is positive, thereby indicating, according to the Clausius–Clapeyron relation \cite{nouchokgwe2025quantifying}, a conventional electrocaloric (EC) effect. In other words, applying an electric field near the transition temperature favors the stabilization of the FE phase, resulting in a positive adiabatic temperature change $\Delta T_{\mathrm{adiab}}$ (see red arrows in Figures \ref{phasediagram}a and \ref{phasediagram}b).

Figure \ref{phasediagram}c presents the phase diagram of PCa2ST, showing a FE–AFE–PE phase sequence. At low temperatures (below 256 K), the material is FE. Between 256 K and room temperature, PCa2ST exhibits an AFE phase, and above room temperature, it becomes PE. Note that the transition temperature of PCa2ST from FE-to-AFE phase at zero electric field was determined by first field-cooling under an applied field of 16 kV\,cm$^{-1}$, followed by heating at zero field (see Supplementary Figure 10), since this transition was not observed in zero-field measurements (see Figure \ref{thermalanalysis}a). The sharpness of the peak observed confirms the first-order character of this transition. The FE-to-AFE transition peak shifts linearly to higher temperatures with increasing electric field, thereby stabilizing the FE phase and yielding a positive $\frac{dT}{dE}$. Consequently, near the FE-to-AFE transition, a conventional EC effect is expected (red arrow in Figure \ref{phasediagram}c). Conversely, the AFE-to-PE transition peak shifts slightly toward lower temperatures, meaning that the electric field stabilizes the PE phase. Here, $\frac{dT}{dE}$ is negative and nearly zero (-0.01\,K\,cm\,kV$^{-1}$). As indicated by the blue arrow in Figure \ref{phasediagram}c, one can therefore anticipate an inverse EC effect ($\Delta T_{\mathrm{adiab}} < 0$ upon application of an electric field) near the AFE-to-PE transition at low electric fields. Under higher electric fields, however, the AFE phase is suppressed and the FE phase becomes stabilized. 

PCa4.6ST exhibits an AFE–PE phase sequence (Figure \ref{phasediagram}d). It remains AFE below 313 K and PE above this temperature. The electric field stabilizes the PE phase, as the AFE-to-PE transition peak shifts toward lower temperatures. A negative $\frac{dT}{dE}$ is observed, thereby indicating an anticipated inverse EC response (blue arrow in Figure \ref{phasediagram}d).

Note that Raman spectroscopy (Supplementary Note 9) carried out under an electric field and temperature confirms the phase diagram measured in PCaxST compounds via in-field calorimetry and \textit{P-E} loops. 

Finally, Figure~3e presents the zero-field phase diagram of highly ordered Ca-doped PST, where the transition temperatures $T$ are shown as a function of calcium content. The non-monotonic behavior of $T$ arises from the fact that calcium stabilizes the AFE phase while destabilizing the FE phase. At low calcium concentrations ($\leq$ 2\% Ca), the FE phase is destabilized, leading to a monotonic decrease in the transition temperature. Specifically, the FE phase is shifted from 297~K (0\% Ca) to 282~K (1\% Ca) and further to 256~K (2\% Ca). At higher calcium contents, the AFE appears and is shifted to higher temperatures. In this regime, the AFE-to-PE peak increases from 297~K (2\% Ca) to 317~K (4.6\% Ca).
Ca-doped PST is AFE for concentrations greater than 1\%, and this AFE phase is more stable with increasing calcium content. The antiferroelectric phase in highly Ca-doped PST was further confirmed microscopically by piezoresponse force microscopy (PFM). The PFM amplitude and phase loops (see Supplementary Note 10) show characteristic antiferroelectric signatures, consistent with previous reports \cite{lu2020probing, chen2017strain}.    

\subsection{First-principles calculations}
In order to gain further understanding of the antiferroelectric behavior observed experimentally, we study the stability of different structural polymorphs of pure and Ca-doped PST using density functional theory (DFT) (see Methods for the computational details). We start by considering an 80-atom supercell of pure PST and restrict ourselves to four structural polymorphs (see Supplementary Figure 25): (i) paraelectric cubic \(Fm\bar{3}m\) (\#225, we denote it further as “c”); (ii) rhombohedral \(R{3}m\) (\#160) which has a polar distortion along a $<$111$>$ direction (we will denote it as “rI” phase); (iii) rhombohedral \(R{3}\) (\#146) which has both a polar distortion along a $<$111$>$ direction and an antiphase oxygen octahedra tilt pattern \textit{a$^{-}$a$^{-}$a$^{-}$} in Glazer notation about the same axis ( “rII” phase); (iv) monoclinic \textit{P}2$_1$/c (\#14) which is derived from the AFE \(Pbam\) phase of PbZrO$_{3}$ by replacing Zr atoms with Sc and Ta cations in the perfectly rocksalt-ordered way; it includes an \textit{a$^{-}$a$^{-}$c$^{0}$} octahedra tilt pattern alongside antipolar displacements of the Pb cations (“AFE” phase). We fully optimize the crystal structures of these polymorphs and compute their energies using DFT. The optimized lattice parameters are listed in Supplementary Table 6, and the computed energies are shown in Supplementary Figure 29a. We find that the rII phase has the lowest energy; however, the AFE polymorph is just 2.7\,meV per A cation higher in energy. A similar result has been reported by Paściak \textit{et al.} for PbSc$_{0.5}$Nb$_{0.5}$O$_{3}$\cite{pasciak2016phase}. Additionally, for PZO, the AFE \(Pbam\) and FE \(R{3}\) phases (where the latter is analogous to the \(R{3}\) phase but with only one type of B cation) were found to be very close in energy, with AFE \(Pbam\) being the ground state \cite{iniguez2014first}. Note, that the addition of zero-point energy contributions does not alter the hierarchy of these polymorphs (see Supplementary Note 11 and Supplementary Table 5 for details). 

Next, we extend this analysis to Ca-doped PST with Ca concentrations of 6.25\% and 12.5\%. For this, we start from the same four polymorphs of pure PST described above (structural polymorhps c, rI, rII and AFE in our 80 atom supercell) and replace one or two Pb atoms, respectively, by Ca atoms as illustrated in Supplementary Figures 26 and 27. We fully optimize all considered structures and compute their energies using DFT. As shown in Supplementary Figure 29a, within the studied range of concentrations, Ca doping does not affect the relative stability of considered polymorphs compared to pure PST. rII remains the lowest energy configuration and the AFE phase is higher in energy by a few meV per A cation. Note, that for 6.25\% Ca-doped PST, the c and rI configurations relax to lower symmetry phases. Namely, the c phase develops the octahedral tilts of the pattern similar to that of the AFE phase, while rI relaxes to the rII phase. In Supplementary Figure 29a we show the energies of these polymorphs with triangular symbols. In turn, the interpolated energy values between 0\% and 12.5\% Ca concentrations are shown with lines.  Structurally, 12.5\% Ca doping leads to a volume reduction of approximately 1\% in the rII phase and 1.3\% in the AFE phase (see Supplementary Table 7 and Supplementary Figure 33). It also increases the amplitude of oxygen octahedral tilts in both rII (by ~14\%) and AFE (by ~19\%) phases, while reducing the polar distortion in the rII phase (by ~6\%) and the antipolar Pb displacements in the AFE phase (by ~37\%).

At the next step, we investigate how the relative stability of rII and AFE polymorphs is affected by other factors, such as hydrostatic pressure and increasing temperature. Since the Ca concentration does not have a significant effect on the stability of these phases in our simulations, we restrict the analysis to the case of pure PST. First, we study how the relative stability of the rII and AFE phases of PST evolves with increasing hydrostatic pressure. For that, we fully optimize the structures of both phases for the pressure values in the range from 0 GPa to 10 GPa (see Supplementary Figure 30 and Supplementary Table 7 for the structural data) and compute the energies of the relaxed structures. As shown in Supplementary Figure 30, the rII phase is favored at low pressures, while the AFE phase becomes more stable above ~3 GPa. At this pressure, the cell volume is reduced by ~3\% in the rII phase and by ~2\% in the AFE phase compared to the zero-pressure structure. In addition, the distortion mode amplitudes are affected as follows: the oxygen octahedral tilts are increased in both rII (by 5.3\%) and AFE (by 7.2\%) phases, polar distortion in the rII phase is reduced by 14.8\%, and antipolar displacements of Pb cations in the AFE phase are reduced by 18.6\%.

Finally, temperature effects are examined by computing the free energies F (within the harmonic approximation as described in Methods) of the rII and AFE phases of PST in the range between 0 K and 500 K. The results (Supplementary Figure 29b) show that the ferroelectric rII phase is favored at low temperatures, and the AFE phase becomes more stable above 170 K. Note that, in our experiments, we did not observe the transition from FE to AFE phase in pure PST. However, in PCa2ST, this transition occurs at approximately 263 K. 

Overall, our first-principles analysis reveals that ferroelectric (rII) and antiferroelectric (AFE) polymorphs of PST are nearly degenerate in energy (at T=0 K), with energy differences of only a few meV per A-site cation, both in pure and Ca-doped systems. Ca substitution up to 12.5\% does not invert the energetic hierarchy, although it induces local structural changes, including enhanced oxygen octahedral tilts and reduced polar (in rII phase) and antipolar (in AFE phase) distortions. In contrast, hydrostatic pressure and finite-temperature effects are found to modify the relative stability of the rII and AFE phases, with pressure above ~3 GPa or temperatures above ~170 K favoring the AFE phase in pure PST. Taken together, these results indicate that the experimentally observed stabilization of the AFE phase in Ca-doped PST likely arises from the combination of having near-degenerate FE and AFE phases, Ca-induced local structural relaxations, and thermal effects, which collectively control the delicate balance between competing instabilities and enable antiferroelectricity.

\subsection{Electrocaloric effect of Ca-doped PST}

\noindent
We next investigate the EC effect of the PCaxST compounds.

From the integration of the zero-field calorimetry peaks (Figure~\ref{thermalanalysis}a), we determine the latent heat $Q$ as a function of calcium content. The results are shown in Supplementary Figure 9. One can see that between 0\% and 2\% Ca, $Q$ drops sharply and then remains nearly constant between 2\% and 4.6\% Ca. For the ferroelectric compounds PST and PCa1ST, we measure latent heats of 932\,J\,kg$^{-1}$ and 544\,J\,kg$^{-1}$, respectively, at the FE-to-PE transition. In PCa2ST, which exhibits two transitions (Supplementary Figure~10 and Figure~\ref{phasediagram}c), we obtain $Q = 109$\,J\,kg$^{-1}$ at the FE-to-AFE transition and $Q = 275$\,J\,kg$^{-1}$ at the AFE-to-PE transition. For the AFE compound PCa4.6ST, we measure $Q = 261$\,J\,kg$^{-1}$. Supplementary Table 1 summarizes the measured entropy change, $\Delta S_0$, associated with thermally driven phase transitions in the different Ca-doped PST compounds. For PST and PCa1ST, thermal driving of the FE–to–PE transition yields $\Delta S_0$ values of 3.1~J~K$^{-1}$~kg$^{-1}$ and 2.1~J~K$^{-1}$~kg$^{-1}$, respectively, indicating a decrease in entropy change with increasing Ca content. In PCa2ST, two successive thermally driven transitions are observed. The FE–to–AFE transition exhibits a $\Delta S_0$ of 0.8~J~K$^{-1}$~kg$^{-1}$, while the AFE–to–PE transition yields a $\Delta S_0$ of 0.9~J~K$^{-1}$~kg$^{-1}$. Finally, PCa4ST undergoes a single thermally driven AFE–to–PE transition, with an associated entropy change of 0.8~J~K$^{-1}$~kg$^{-1}$.  

\begin{figure}[!ht]
    \centering
    \vspace{3ex}%
    \includegraphics[width=1.1\textwidth]{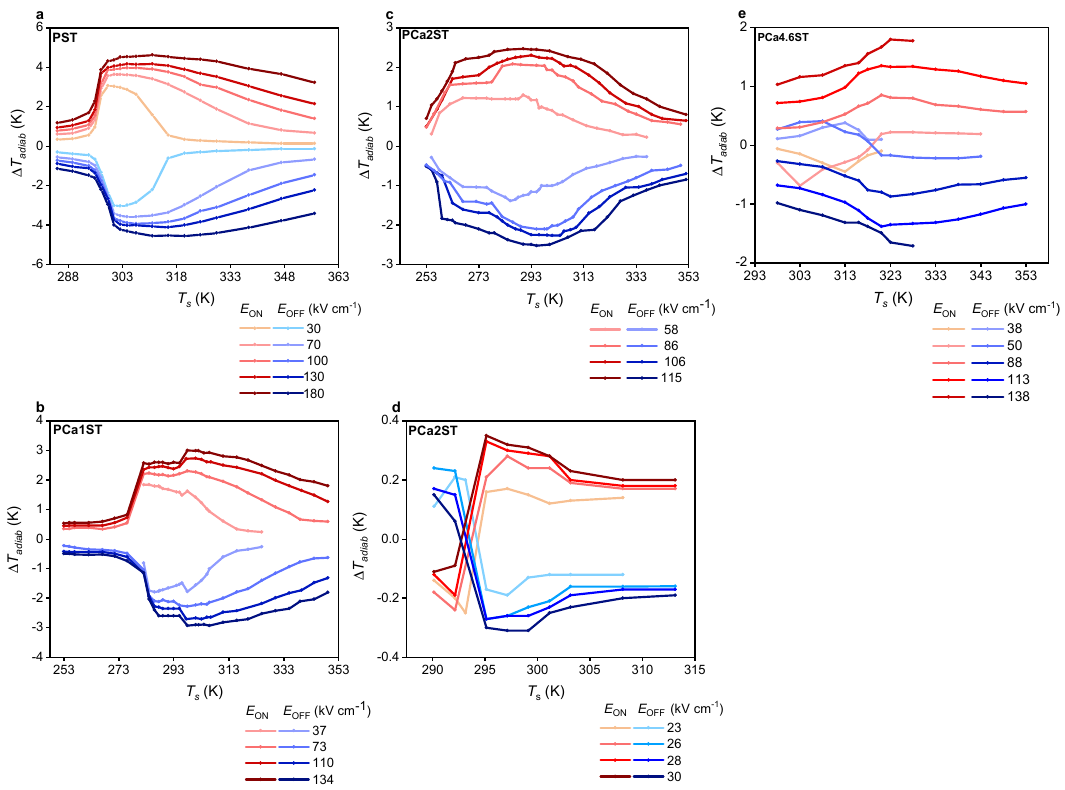}
    \caption{\textbf{Electrocaloric effect of Ca-doped PST.} Adiabatic temperature change, $\Delta T_{\mathrm{adiab}}$, of the different Ca-doped compositions measured at various starting temperatures ($T_s$) and electric fields ($E$). A conventional EC effect is observed in PST (a) and PCa1ST (b). In PCa2ST, a conventional EC effect is obtained at high electric fields ($> 30$ kV cm$^{-1}$) (c), whereas an inverse EC effect is observed near the AFE–PE transition at low electric fields ($\leq 30$ kV cm$^{-1}$) (d). PCa4.6ST exhibits an inverse electrocaloric effect (e).}  
    \label{ECE}
\end{figure}

We use an infrared (IR) camera to estimate the adiabatic temperature change $\Delta T_{\mathrm{adiab}}$ of the compounds at different starting temperatures and electric fields (see Figure \ref{ECE} and Supplementary Table 3).

As expected from its phase diagram (Figure \ref{phasediagram}a), PST exhibits a conventional EC effect (Figure \ref{ECE}a), characterized by heating upon field application and cooling upon field removal. Both $\Delta T_{\mathrm{adiab}}$ and its temperature range increase with electric field. We measure a maximum $\Delta T_{\mathrm{adiab}}$ of 4.6\,K at 316\,K under an electric field of 180\,kV\,cm$^{-1}$. The maximum value occurs slightly above the Curie temperature, as the peak shifts toward higher temperatures with increasing field (see Figure~\ref{phasediagram}a). This $\Delta T_{\mathrm{adiab}}$ is comparable to the highest value reported \cite{Nair2019} for PST multilayer capacitors (MLCs), where larger electric fields can be achieved due to the multilayer geometry.

In PCa1ST, at its FE-to-PE transition (282\,K), we also observe a conventional EC effect that occurs 17\,K lower than in undoped PST. The amplitude and temperature range of the EC effect in PCa1ST broaden with increasing field. We measure a maximum $\Delta T_{\mathrm{adiab}}$ of 2.6\,K at 285\,K under 134\,kV\,cm$^{-1}$, which remains nearly constant up to 298 K and then increases to 3 K at this temperature.

We further examine the EC response of PCa2ST in detail. PCa2ST exhibits both conventional (Figure~\ref{ECE}c) and inverse (Figure~\ref{ECE}d) electrocaloric effects, depending on the applied electric field and temperature.
A conventional EC effect is observed upon application of high electric fields ($> 30$~kV\,cm$^{-1}$) (see Figure~\ref{ECE}c). At $115$~kV\,cm$^{-1}$, the maximum $\Delta T_{\mathrm{adiab}}$ reach $2.0$~K and $2.5$~K at the FE--to--AFE ($264$~K) and AFE--to--PE ($297$~K) phase transitions, respectively. The conventional EC response at the FE--to--AFE transition originates from the positive slope of $\mathrm{d}T/\mathrm{d}E$ (Figure~\ref{phasediagram}c). For the AFE--to--PE transition, when the electric field exceeds $30$~kV\,cm$^{-1}$, the material is driven into the FE phase (see Figure \ref{phasediagram}c). Since the FE phase has a lower entropy than the AFE phase, the field-induced transition results in a conventional EC effect.
In contrast, the inverse EC effect in PCa2ST (Figure~\ref{ECE}d) appears near the AFE--to--PE phase transition at low electric fields ($\leq 30$~kV\,cm$^{-1}$). Below $295$~K, a persistent but low-magnitude inverse EC response is observed. The maximum inverse $\Delta T_{\mathrm{adiab}}$ reaches $-0.25$~K, with the peak shifting toward lower temperatures as the electric field increases. This behavior is consistent with the PCa2ST phase diagram, which shows a negative $\mathrm{d}T/\mathrm{d}E$ slope at the AFE--to--PE transition. Moreover, the magnitude of the inverse EC effect decreases with increasing electric field and transitions to a conventional response above $30$~kV\,cm$^{-1}$ (see Supplementary Note~7). Above $295$~K, the EC response becomes entirely conventional, as the material is stabilized in the paraelectric (PE) phase.

Finally, we investigate the EC effect in the antiferroelectric material PCa4.6ST. As shown in Figure~\ref{ECE}e, an inverse EC effect is observed in the vicinity of the AFE–to–PE transition. At 313~K under an applied electric field of 38~kV\,cm$^{-1}$, an inverse EC temperature change of $-0.5$~K is measured. With increasing electric field, the EC peak shifts toward lower temperatures, reaching a maximum inverse temperature change of $-0.6$~K at 303~K under 50~kV\,cm$^{-1}$.
At higher temperatures and electric fields, the EC response transitions to a conventional effect. A maximum conventional EC temperature change of 1.8~K is recorded at 323~K under high electric field (Figure~\ref{ECE}e). As indicated in Figure~\ref{phasediagram}d, an inverse EC response is expected in this region because $\mathrm{d}T/\mathrm{d}E$ is negative. Moreover, the \textit{P-E} loops of PCa4.6ST (Figure~\ref{peloops}d and Supplementary Figure~6) reveal an increase in polarization with temperature, corresponding to a positive $\partial P/\partial T$ and therefore a negative $\Delta T_{\mathrm{adiab}}$, consistent with the Maxwell relation~\cite{Moya2014}. At sufficiently high electric fields and temperatures, the paraelectric phase becomes energetically favored, resulting in a positive (conventional) EC effect.

\section{Discussion}

By doping PST on the A-site with calcium, we modify its phase transitions and tune its electrocaloric response. The Ca-doped PST compounds retain a high degree of B-site cation ordering, comparable to that of undoped PST. We detect and measure antiferroelectricity in Ca-doped PST for concentrations greater than 1\%. Specifically, PST and PCa1ST exhibit a FE-to-PE transition, PCa2ST shows an FE-to-AFE-to-PE transition, and PCa4.6ST displays an AFE-to-PE transition. Calcium stabilizes the AFE phase and destabilizes the FE phase. It is worth noting that the degree of B-site cation ordering plays a crucial role. In PST, a high level of disorder leads to relaxor behavior \cite{yin2025optimizing}, which may hinder the stabilization of a well-defined antiferroelectric phase upon Ca doping. 

An antiferroelectric phase between the ferroelectric and paraelectric phases as observed in PCa2ST, has also been reported in other lead-based perovskite materials \cite{Tan2010}. The archetypal AFE material, PbZrO$_3$ (short form - PZO) \cite{Tagantsev2013}, when doped on the A-site with Nb \cite{Novak2018} or La \cite{pan1989field} and on the B-site with Sn and Ti, exhibits a sequence of phase transitions from FE to AFE to PE, which contrasts with the more typical phase sequence (AFE to FE to PE) observed for example in pure PZO \cite{Pablo2021}. In PZO, an electric field induces a stable FE phase between the AFE and PE states, and the temperature range of this FE phase broadens with increasing electric field \cite{Pablo2021}. In contrast, in PCa2ST, the electric field stabilizes the FE phase at lower temperatures while progressively suppressing the AFE region as the field strength increases (see Figure \ref{phasediagram}c). 

Given the significant difference in ionic radii between Pb (1.49\,Å) and Ca (1.12\,Å) \cite{Shannon1976}, it is natural to expect that chemical pressure due to Ca doping will affect the phase transition in Ca-doped PST like hydrostatic pressure. However, the appearance of the AFE phase in Ca-doped PST cannot be explained solely by lattice contraction due to Ca substitution. Our DFT calculations (see Sec 2.5 and Supplementary Note 11) show that the FE rII and AFE polymorphs are nearly degenerate (in the limit of the accuracy of DFT; see e.g. \cite{aramberri2021possibility}) at $T=0$~K in both pure and Ca-doped PST, indicating that even small perturbations can shift the balance between the two phases. While Ca doping indeed reduces lattice constants and cell volumes (12.5\% doping equivalent to an effective pressure of $\sim$1--2~GPa), this alone is insufficient to stabilize the AFE phase, which requires $\sim$3~GPa in pure PST (see Sec. 2.5 and Supplementary Note 11). Analysis of symmetry-adapted distortion modes (Supplementary Note 11) reveals that Ca doping enhances octahedral tilts in both rII and AFE phases, while reducing the antipolar Pb displacements in the AFE phase and the polar distortion in the rII phase. Consequently, our analysis suggests that the experimentally observed AFE ordering in Ca-doped PST (above $\sim$1\% Ca) arises from the combined effects of near-degenerate of FE and AFE phases, local distortions around Ca ions, and temperature-driven stabilization, rather than from volume reduction alone.

The ability of calcium to influence the Curie temperature of PST while maintaining a significant electrocaloric response is of particular interest for electrocaloric cooling. Having little Ca content ($\leq$2\%Ca) lowers the transition temperature, and higher level calcium content ($\geq$2\%Ca) shifts it to higher temperatures. Ca-doped PST could enable the design of a layered regenerator capable of operating between 263\,K and 353\,K. Layered regenerators, commonly used in magnetocaloric cooling \cite{NAVICKAITE2018}, combine materials with different Curie temperatures to achieve an extended operational temperature span. Shebanov et al. \cite{Shebanov1992} previously demonstrated that the transition temperature of PST can be tuned by doping the B-site with specific ions, although A-site calcium substitution was not considered in their study. The authors reported that the transition temperature could be reduced to 257\,K, albeit with a negligible electrocaloric effect of only 0.25 K. 

\begin{figure}[!ht]
    \centering
    \vspace{3ex}%
    \includegraphics[width=1\textwidth]{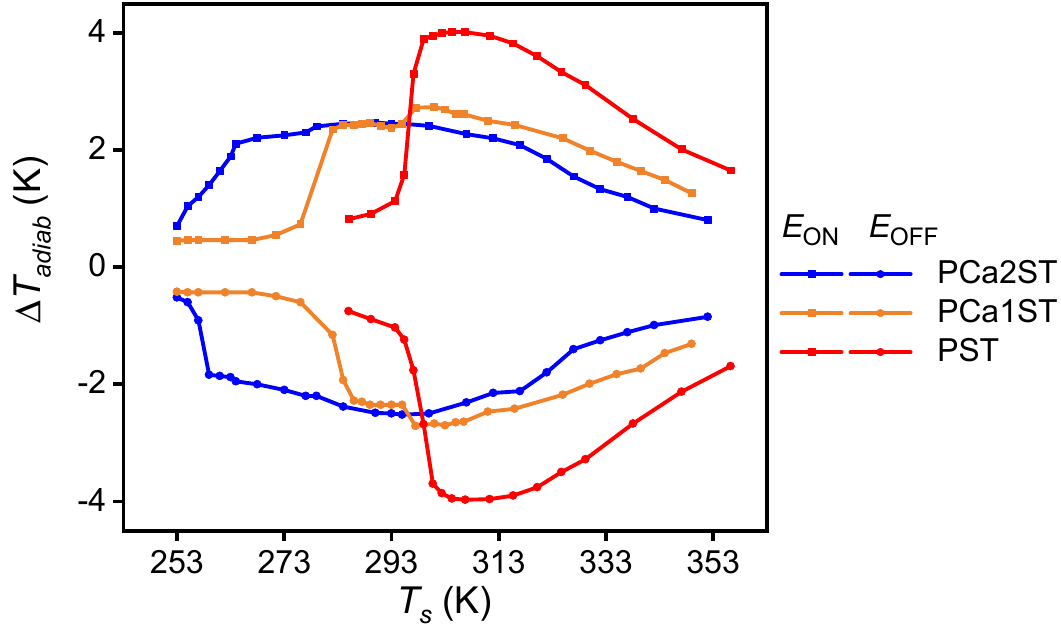}
    \caption{\textbf{Controlling the electrocaloric effect through A-site doping.} Adiabatic temperature change ($\Delta T_{\mathrm{adiab}}$) of PST, PCa1ST, and PCa2ST as a function of the starting temperature ($T_s$), measured under the same applied electric field of 110\,kV\,cm$^{-1}$.}
    \label{controlECE}
\end{figure}

Figure~\ref{controlECE} illustrates how the conventional EC effect of Ca-doped PST (undoped PST, PCa1ST, and PCa2ST) can be exploited to extend the operating temperature range of PST toward lower temperatures. A maximum adiabatic temperature change of 1.8\,K in PST, observed between 293\,K and 313\,K, has been implemented in a state-of-the-art electrocaloric cooling device \cite{li2023high}. Based on this benchmark, we adopt $\Delta T_{\mathrm{adiab}}$ $\geq$ 1.8\,K as a threshold criterion to define the practical operating temperature range. Undoped PST operates between 295\,K and 353\,K with $\Delta T_{\mathrm{adiab}} \geq$ 1.8\,K. Upon Ca doping, the operating window shifts toward lower temperatures: PCa1ST (1\% Ca) operates between 280\,K and 336\,K, while PCa2ST (2\% Ca) covers the range from 263\,K to 323\,K. In an EC regenerator, combining PCa2ST with PST would enable coverage of a broad temperature range from 263\,K to 353\,K while maintaining $\Delta T_{\mathrm{adiab}} \geq$ 1.8\,K. For comparison, the EC effect reported for less ordered bulk PST \cite{yin2025optimizing} reaches approximately 1.8\,K over a temperature interval of 288–318\,K. The 4.6\% Ca-doped composition is not considered here, as its small negative temperature change under applied electric field would counteract the positive EC response of the other compositions. In future work, the fabrication of multilayer capacitors \cite{hirose2016progress} using Ca-doped PST could enable the application of higher electric fields and potentially further expand the material’s operational temperature range as demonstrated for pure PST in \cite{Nair2019}.

In conclusion, we have demonstrated that A-site calcium doping in highly ordered lead scandium tantalate enables a substantial electrocaloric effect at temperatures below the freezing point of water. Ca-doped PST compositions cover a temperature range from 263\,K to 353\,K with an electrocaloric response of 2\,K. Furthermore, with increasing Ca content, an intermediate antiferroelectric phase emerges between the ferroelectric and paraelectric states, giving rise to an inverse electrocaloric effect at high Ca concentration. These findings are supported by a combination of Raman spectroscopy, calorimetry, electrical polarization measurements, piezoresponse force microscopy, and density functional theory calculations. Overall, our results highlight the potential of A-site doping as an effective strategy to broaden the operational temperature range of PST. Ca-doped PST thus represents a promising candidate for use in cascaded electrocaloric cooling devices. 

\section*{Methods}

\subsection*{Sample preparation:} 
PbO (product no. 12220, Alfa Aesar, USA), Sc$_2$2O$_3$ (product no. 11216, Alfa Aesar, USA), Ta$_2$O$_5$ (product no. 73118, Alfa Aesar, USA) and CaCO$_3$ (product no. 10996, Alfa Aesar, USA) were used to synthesize the PCaxST powders. The homogenized, stoichiometric mixture was mechanochemically activated in a high-energy planetary mill (model PM 400, Retsch, Germany) with 15 WC balls (2r = 20 mm) in a 250\,mL WC vessel for 24 h at 300\,rpm. The powder was then heated at 1000 °C for 2 h in a closed alumina crucible. The synthesized powder was milled in a planetary mill with yttria-stabilized zirconia balls in isopropanol at 200 rpm for 2 h to achieve deagglomeration. The powder was then uniaxially pressed into cylindrical pellets and further consolidated by isostatic pressing at 300 MPa (Autoclave Engineers, USA). The powder compacts were sintered at 1300 °C for 2 hours and annealed at 1000 °C for 48 hours to achieve a high B-site ion ordering in the pellets. This thermal treatment of the pellets was carried out in the packing powder, which consisted of PbZrO$_3$ with an addition of 2 mol\% PbO powder in excess. The density of the ceramics was measured using a gas displacement density analyzer (AccuPyc II1340 Pycnometer, Micromeritics, USA). The theoretical density value was determined from line card no. 00-043-0134 of the powder sample analysis program X'Pert HighScore Plus.

\subsection*{Scanning electronic microscope:}
The microstructure of the sintered pellets was examined by a scanning electron microscope (JEOL JSM-7600F, Jeol Ltd., Japan) at 15\,kV in backscattered electron mode (see Supplementary Note 1). The samples were ground and fine-polished by a conventional metallographic procedure and thermally etched at 950\,°C for 5 to 10\,min. Grain size distribution (Supplementary Figure 2) was determined by analyzing digitalized microstructure images in Image Tool software (version 3.0, United States). 

\subsection*{X-ray diffraction:}
XRD of the crushed PCaxST ceramic samples was carried out with the PANalytical X’Pert PRO (Malvern Panalytical, Malvern, UK) using Cu-K$_{\alpha1}$ radiation. The diffraction patterns were measured in the 2$\theta$ range of 15–60° with a step of 0.039° and a dwell time of 100\,s per step. The B-site cation ordering $\Omega$ was computed using the equation below defined in \cite{Shebanov1989}.
\begin{equation}
    \Omega^2 = \frac{(\frac{I_{111}}{I_{200}})_{exp}}{(\frac{I_{111}}{I_{200}})_{theory, \Omega = 1}}
    \label{eq:order}
\end{equation}
\\
$I_{111}$ and $I_{200}$ represent the integrated intensities of the pseudocubic perovskite diffraction peak 111 and 200 respectively. The experimental XRD patterns of the four samples are presented in Supplementary Figure 3.

The theoretical XRD patterns of pure and Ca-doped PST are computed using the VESTA package \cite{aramberri2021possibility}. In all cases, we consider the cubic structures with perfect order of Sc and Ta cations. For pure PST, we compute the XRD pattern using 40 atom cell (2x2x2 of the 5 atom perovskite cell). The result is shown in Supplementary Figure 28. Then, we compute the ratio of the integrated intensities ($\frac{I_{111}}{I_{200}})_{theory}$  of (111) and (200) peaks. The resulting value is 1.37, which is in agreement with previously reported 1.33 \cite{wang1990order} and 1.4 \cite{Shebanov1989}. For Ca-doped PST, we compute the XRD patterns using 320 atom cell (4x4x4 of the 5 atom perovskite cell). We consider three Ca concentrations: 1.56\%, 3.125\% and 12.5\%, which correspond to replacing 1, 2 and 8 Pb atoms (out of 64) by Ca, respectively. For 3.125\% Ca content, we create 10 configurations with 2 Ca atoms replacing randomly selected Pb atoms. Then, we compute corresponding XRD patterns and extract the ($\frac{I_{111}}{I_{200}})_{theory}$ ratios. The averaged ratio over 10 configurations is 1.47. The same analysis is repeated for 12.5\% Ca doping with the resulting averaged ratio of 2.598. Finally, the ($\frac{I_{111}}{I_{200}})_{theory}$ ratios for PST doped with 1\%, 2\%, 3\% and 4.6\% Ca are obtained by interpolation between the values corresponding to 0, 1.56, 3.125 and 12.5\% (see the results in Supplementary Table 4).  

\subsection*{Polarization versus electric field loops:}
The isothermal electrical polarization loops of the four ceramic samples were collected using an AixACCT TF Analyzer 2000 at a frequency of 1 Hz. Gold electrodes were deposited on both the top and bottom surfaces of the samples, and silver paste was used to attach wires to the electrodes. The surface electroded correspond to an area of 19.6\,mm$^2$. A maximum voltage of 1100\,V was applied to the PST, PCa1ST, PCa2ST, and PCa3ST samples, while up to 2000\,V was applied to the PCa4.6ST sample.

\subsection*{Differential scanning calorimetry:}
Using a customized Differential Scanning Calorimeter (DSC) Mettler Toledo 3+, heat flow ($dQ/dT$) measurements were performed on the four ceramic compositions at a heating/cooling rate of 5\,K/min. After baseline correction, the $dQ/dT$ peaks were integrated using the trapezoid method to calculate the latent heat ($Q$) during both cooling and heating cycles. Isofield measurements—i.e., heat flux measurements under a constant electric field while varying the temperature—were also conducted on the same samples using the customized DSC to construct phase diagrams ($T_c$ = f($E$)). Voltages ranging from 0 to 600\,V were applied using a Keithley 2410 sourcemeter.

\subsection*{Raman spectroscopy:}
The Raman spectroscopic measurements were conducted with Renishaw’s InVia Raman microscope in backscattering configuration, with a 20X  long-distance objective. A 532 nm (2.33 eV) laser was used for excitation, with an optimized laser power of 4.85 mW to minimize the laser induced heating effect. The electric field was applied with a Keithley 2470 source meter via a top electrode of ITO (100nm) and a bottom electrode of gold. Sample temperatures above 283 K were controlled using a Linkam THMS600 temperature controller, while measurements below 283 K were performed with a cryostat from Oxford Instruments, both of which were equipped with internal electrical contacts. The measurements were carried out on heating.

\subsection*{Dielectric spectroscopy:}
The temperature-dependent dielectric permittivity $\varepsilon’$ was measured using a Novocontrol Concept 40 dielectric spectrometer system. Conductive silver paint was used to fix the electroded bulk samples in the sample holder. Starting from room temperature, the temperature was repeatedly cycled between lower and upper boundaries depending on the sample composition with a heating/cooling rate of less than 1 K/min. Permittivity was measured at frequencies between 1\,MHz and 1\,kHz with a voltage amplitude of 0.5 V\,rms.

\subsection*{Density functional theory:} 
All first-principles calculations were performed using the Vienna \textit{ab initio} Simulation Package (VASP) \cite{kresse1996efficient} within the projector-augmented plane wave method \cite{blochl1994projector} of density functional theory \cite{hohenberg1964density, kohn1965self}. The following electronic orbitals were treated explicitly: 6\textit{s} and 6\textit{p} for Pb; 3\textit{s}, 3\textit{p} and 4\textit{s} for Ca; 3\textit{d} and 4\textit{s} for Sc; 5\textit{d} and 6\textit{s} for Ta; and 2\textit{s} and 2\textit{p} for O. We employed the generalized gradient approximation for the exchange-correlation functional in the form of Perdew, Burke, and Ernzerhof optimized for solids (PBEsol) \cite{perdew2008restoring}. The cutoff energy for the plane-wave basis set is set to 500\,eV. We used 3x3x3 $\Gamma$-centered Monkhorst-Pack k-point grid for reciprocal space integrals in the Brillouin zone corresponding to 80 atom supercell. The electronic densities of states for pure and Ca-doped PST were computed using 9x9x9 k-point grid. We ensured that these parameters provide well-converged results for the quantities of interest. In the lattice optimizations, the structures were considered as relaxed if the forces acting on the atoms are below 0.005\( \text{eV/\AA} \)  and stresses are below 0.1\,kBar.  

\subsection*{Symmetry analysis:}
The symmetry analysis of the crystal structures was performed using FINDSYM tool from ISOTROPY package \cite{stokes_isotropy, stokes2005findsym}. 

\subsection*{Zero-point energy and free energy calculations:}

We employed VASP and PHONOPY \cite{togo2023, togo2023implementation} to compute phonons using finite-displacement supercell approach. We utilized 2x2x2 supercells with different sets of atomic displacements constructed from the 80 atom rII and AFE cells of PST. We used 2x2x2 $\Gamma$-centered Monkhorst-Pack k-point grid in the calculation of forces. Phononic integrals were performed using 8x8x8 $\Gamma$-centered k-point grid. This provides the convergence of zero-point energy values within 0.1\,meV. The non-analytical correction was included in all calculations. The k-path for computing phonon bands was generated using SeeK-path tool \cite{hinuma2017band}.

Zero-point energies were calculated using the following expression \cite{aramberri2021possibility, phonopy}:
\begin{equation}
E_{\text{ZP}} = \frac{1}{2} \sum_{q\nu} \hbar \omega_{q\nu},
\end{equation}
where \(\hbar\) is the reduced Planck constant, and \(\omega_{q\nu}\) is the frequency of the phonon with wavevector \(q\) and branch index \(\nu\).

Free energies were calculated as follows \cite{aramberri2021possibility, phonopy}:
\begin{equation}
F = E_{\text{KS}} + E_{\text{vib}} - TS.
\end{equation}
Here, \(E_{\text{KS}}\) is the Kohn-Sham energy obtained using DFT, and \(E_{\text{vib}}\) is the vibrational energy defined as:
\begin{equation}
E_{\text{vib}} = \sum_{q\nu} \left[ \frac{1}{2} + n_{q\nu} \right] \hbar \omega_{q\nu} 
= E_{\text{ZP}} + \sum_{q\nu} \hbar \omega_{q\nu} n_{q\nu},
\end{equation}
where \(n_{q\nu}\) is the number of phonons with a given \(q\) and \(\nu\):
\begin{equation}
n_{q\nu} = \frac{1}{\exp{\left(\frac{\hbar \omega_{q\nu}}{k_B T}\right)} - 1}.
\end{equation}
Here, \(k_B\) is the Boltzmann constant, and \(T\) is the temperature.

The entropy \(S\) is defined as:
\begin{equation}
S = -k_B \sum_{q\nu} \ln \left[ 1 - \exp \left( -\frac{\hbar \omega_{q\nu}}{k_B T} \right) \right] 
+ \frac{1}{T} \sum_{q\nu} \hbar \omega_{q\nu} n_{q\nu}.
\end{equation}

Therefore, the final expression for the free energy can be written as:
\begin{equation}
F = E_{\text{KS}} + E_{\text{ZP}} + k_B T \sum_{q\nu} \ln \left[ 1 - \exp \left( -\frac{\hbar \omega_{q\nu}}{k_B T} \right) \right].
\end{equation}

\subsection*{Crystal structure visualization and theoretical XRD patterns:}
The VESTA package \cite{momma2011vesta} was used to visualize crystal structures shown in Supplementary Note 11 and to compute the theoretical XRD patterns of perfectly ordered pure and Ca-doped PST. 

\subsection*{Piezoresponse force microscopy (PFM) measurements:}
Prior to the PFM measurements, the samples were cut and polished using standard metallographic techniques. The thickness of the samples was \SI{150}{\micro\meter}. The PFM images were acquired using an atomic force microscope (AFM; Asylum Research, MFP-3D, Santa Barbara, CA, USA) equipped with a high-voltage PFM module. A tetrahedral silicon tip coated with Ti/Ir (ASYELEC.01-R2, Oxford Instruments, Germany) with a tip radius of \SI{25}{\nano\meter} was used. The spring constant $k$ and resonance frequency $f$ of the cantilever were \SI{2.8}{\newton\per\meter} and \SI{75}{\kilo\hertz}, respectively. The out-of-plane PFM images were recorded in dual AC resonance tracking (DART) mode by applying an AC voltage with an amplitude of \SI{7}{\volt} and a frequency of approximately \SI{320}{\kilo\hertz} between the PFM tip and the bottom electrode (Ted Pella Silver Paste, product number 16035, Redding, California, USA).

The local PFM amplitude and phase hysteresis loops were measured in the PFM switching spectroscopy (SS) mode with the pulsed DC step signal and the superimposed AC drive signal as described in \cite{urvsivc2019investigations}. The measurements were performed in the on-electric-field SS mode and off-electric-field SS mode for Ca-doped PST and PST samples, respectively. The waveform parameters were: the sequence of the rising DC step signal was driven at \SI{20}{\hertz} with a maximum amplitude of \SI{7}{\volt}; the frequency of the triangular envelope was \SI{0.2}{\hertz}; a superimposed sinusoidal AC signal with an amplitude of \SI{2}{\volt} and a frequency of \SI{320}{\kilo\hertz} was used. One cycle was recorded in an on-electric field mode and off-electric-field mode.

An \textit{in-situ} electrical poling experiment was performed on the Ca-doped PST sample. First, the \SI{10}{\micro\meter} $\times$ \SI{10}{\micro\meter} PFM image was scanned in DART mode with an AC voltage of \SI{7}{\volt} and frequency of \SI{340}{\kilo\hertz}. After that, the \SI{5}{\micro\meter} $\times$ \SI{5}{\micro\meter} area was scanned by a DC electric field with an amplitude of \SI{200}{\volt} (\textit{in-situ} poling). In the third step, the same \SI{10}{\micro\meter} $\times$ \SI{10}{\micro\meter} area as at the beginning was scanned again with an AC voltage of \SI{7}{\volt}, immediately after \textit{in-situ} DC poling ($t \approx 0$~min) and after 20 hours. All PFM measurements were performed at a temperature of \SI{27}{\celsius}.

\subsection*{Infrared camera}
Direct measurements of the EC effect were performed using an infrared (IR) camera (A6751sc, FLIR) to determine the adiabatic temperature change $\Delta T_{\mathrm{adiab}}$ of the samples. The temperature acquisition frequency was set to 900\,Hz above 10\,$^{\circ}$C and 500\,Hz between –20\,$^{\circ}$C and 10\,$^{\circ}$C. The Ca-doped PST samples were electroded with gold on both surfaces (see Sample Preparation in the Methods section). Silver paste was used to attach electrical leads to the top and bottom electrodes. To enhance emissivity, the top surface of each sample was coated with black paint, achieving an emissivity close to 1. The sample temperature was controlled using a Linkam cell (THMS600), which was sealed with a barium fluoride (BaF$2$) window to allow infrared transmission while preventing moisture condensation at low temperatures. Three electric-field cycles were applied at different starting temperatures, and the average of the three measurements was used to determine $\Delta T_{\mathrm{adiab}}$. The experiments were conducted over a temperature range from 253\,K to 353\,K. A Keithley 2410 source meter was used to apply voltage under constant-current conditions to the Ca-doped compounds. To maintain adiabatic conditions, a current of 1\,mA was used. An automated Python-based control system synchronized the IR camera, Linkam stage, and Keithley source meter to enable simultaneous and rapid data acquisition.

\subsection*{Thermistor}
We measured the EC effect of PCa4.6ST samples directly using a small-bead thermistor glued on the sample surface and connected to a digital multimeter. For the measurements, a square electric-field waveform with a magnitude of 10--100~kV\,cm$^{-1}$ was used. The samples were placed inside a modified differential scanning calorimeter (Netzsch DSC~204~F1, Germany), which enables highly precise temperature stabilization within $\pm$0.001~°C. The reported $\Delta T_\mathrm{adiab}$ values correspond to the measured maximum temperature changes multiplied by a correction factor of 1.8. This factor compensates for thermal losses in the non-active components of the system, including the wires, thermistor, electrodes, adhesive, and the passive (non-electroded) parts of the sample. Details of the method can be found in~\cite{kutnjak2013indirect}.

\section*{Data availability}
All relevant data are available on request from the corresponding authors. Source data are provided with this paper.

\section*{Acknowledgements:}
Y.N., N.S.F., V.K., U.P., T.G., J.I.G. and E.D. acknowledge the Fonds National de la Recherche (FNR) of Luxembourg through the THERMODIMAT project C20 / MS / 14718071 /Defay. I.G., N.S., S.D., M.S., and H.U. are grateful to the Slovenian research agency grants (No. J2-60035, N2-0212, No. P2-0105, and the Young Investigator Program). B.M. and J.I.G. are thankful to the Luxembourg National Research Fund through grant INTER/NOW/20/15079143/ TRICOLOR.

\section*{Author contributions}
E.D. designed and supervised the project. Y.N. conducted the experiments that led to Figures 1–5 and Supplementary Notes 3 and 5–8. I.G., S.D., and H.U. prepared the Ca-doped PST samples and, with assistance from H.U., carried out the experiments that led to Supplementary Notes 1 and 2. N.S. acquired the PFM and AFM data, with assistance from H.U., leading to Supplementary Note 10. M.S. performed the experiments that led to Supplementary Figure 16. T.G. collected the dielectric measurements that led to Supplementary Note 4. M.M. deposited interdigitated electrodes on the samples for Raman measurements. Y.N. and P.B., with assistance from G.F.N. and M.G., collected the Raman spectroscopy data that led to Supplementary Note 9. N.S.F., with assistance from B.M. and J.I.G., performed the density functional theory calculations that led to Supplementary Note 11. Y.N. wrote the manuscript with contributions from N.S.F. and E.D.; V.K., I.G., N.S., M.S., U.P., G.F.N., T.G., M.G., H.U. and J.I.G. contributed to the final version of the manuscript. Y.N. wrote the supplementary information with contributions from N.S.F. and M.G., and feedback from N.S., T.G., and E.D.; E.D. obtained the funding.
\section*{Competing interests}
The authors declare no competing interests.

\clearpage

\begin{table}[ht]
\centering
\caption{\textbf{B-site cation ordering of PCaxST.} This table describes the density, average grain size and B-site cation ordering ($\Omega$) of the four ceramics studied. $I_{111}$ and $I_{200}$ are respectively integrated intensities of XRD maxima 111 and 200. The experimental and simulated XRD plots are shown in Supplementary Figure 3 and Supplementary Figure 28, respectively. Theoretical $I_{111}$/$I_{200}$ ratios are computed as described in the Methods section. $\Omega$ is calculated using the equation defined in the Methods section.}
\begin{tabularx}{\textwidth}{|>{\centering\arraybackslash}X|>{\centering\arraybackslash}X|>{\centering\arraybackslash}X|>{\centering\arraybackslash}X|>{\centering\arraybackslash}X|>{\centering\arraybackslash}X|}
\hline
\textbf{Sample} & \textbf{Density (kg/m\textsuperscript{3})} & \textbf{Average grain size ($\mu$m)} & \textbf{Theo. $I_{111}/I_{200}$} & \textbf{Exp. $I_{111}/I_{200}$} & \textbf{$\Omega$} \\
\hline
PST       & 8760  & 1.22 & 1.37 & 1.07 & 0.88 \\
\hline
PCa1ST    & 8650  & 1.45 & 1.43 & 1.22 & 0.92 \\
\hline
PCa2ST    & 8740  & 1.26 & 1.50 & 1.25 & 0.91 \\
\hline
PCa4.6ST  & 8640  & 1.65 & 1.75 & 1.37 & 0.89 \\
\hline
\end{tabularx}
\label{tablePCaxST}
\end{table}

\clearpage

\clearpage

\end{document}